# State Transfer in Latent-Symmetric Networks

Jonas Himmel[1], Max Ehrhardt[1], Matthias Heinrich[1], Sebastian Weidemann[1], Tom A. W. Wolterink[1], Malte Röntgen[2,3], Peter Schmelcher[2,4], and Alexander Szameit[1]

## Abstract

The transport of quantum states is a crucial aspect of information processing systems, facilitating operations such as quantum key distribution and inter-component communication within quantum computers. Most quantum networks rely on symmetries to achieve an efficient state transfer. A straightforward way to design such networks is to use spatial symmetries, which severely limits the design space. Our work takes a novel approach to designing photonic networks that do not exhibit any conventional spatial symmetries, yet nevertheless support an efficient transfer of quantum states. Paradoxically, while a perfect transfer efficiency is technically unattainable in these networks, a fidelity arbitrarily close to unity is always reached within a finite time of evolution. Key to this approach are so-called latent, or 'hidden', symmetries, which are embodied in the spectral properties of the network. Latent symmetries substantially expand the design space of quantum networks and hold significant potential for applications in quantum cryptography and secure state transfer. We experimentally realize such a nine-site latent-symmetric network and successfully observe state transfer between two sites with a measured fidelity of 75%. Furthermore, by launching a two-photon state, we show that quantum interference is preserved by the network. This demonstrates that the latent symmetries enable efficient quantum state transfer, while offering greater flexibility in designing quantum networks.

## Main

The search for symmetries is commonly regarded as the guiding principle for unveiling the laws of nature. Through Noether's theorem, they are the basis for all known conservation laws, such as the conservation of energy, momentum and charge. Symmetries in real space include translation and inversion ('parity') symmetry, as occurring in the position space of crystalline matter, or permutation symmetry found in finite networks (Fig. 1a). The presence of translation symmetry gives rise to band structures in momentum space, which form the very foundation of solid-state physics (Fig. 1b). Remarkably, any system in momentum space can be fully classified by ten combinations of only three fundamental symmetries, particle-hole, chiral and time-reversal, in the ten-fold way[1,2]. These fundamental symmetries show up in the band structures as well as the associated eigenstates, and may establish protected topological phases, such as in topological insulators and superconductors[3,4].

Real-space symmetries like permutation symmetry apply also to finite arrangements, where they influence e.g. the dynamics in quantum spin networks. Such networks play a crucial role in quantum information processing, particularly in the generation[5,6], fractional revival[7,8], and transfer[9,10] of states. Designing networks capable of carrying out those tasks is greatly aided by symmetries between certain sites of the network. Conventionally, two network sites are considered symmetric if their permutation preserves the overall structure of the network (Fig 1a, right). Recently, so-called latent or 'hidden' symmetries[11–14] have been theoretically introduced, which instead are embodied in the spectral properties of the network (Fig. 1c) without leaving obvious signatures in either real- or momentum-space. This novel concept of symmetry also influences the networks dynamics and has developed a deeper understanding of intriguing phenomena like degeneracies[15], flat bands[16], topology[17–19] and state transfer[20–22]. So far, despite the promising potential for quantum state transfer, an experimental implementation of dynamics in latent symmetric networks remains elusive.

[1] Institute for Physics, University of Rostock, Rostock, Germany.
[2] Center for Optical Quantum Technologies, University of Hamburg, Hamburg, Germany.
[3] Eastern Institute for Advanced Study, Eastern Institute of Technology, Ningbo, Zhejiang 315200, People's Republic of China.
[4] The Hamburg Centre for Ultrafast Imaging, University of Hamburg, Hamburg, Germany.

In this work, we experimentally realize an integrated photonic network that contains two latent-symmetric network sites without exhibiting any conventional symmetry. We study the dynamics of light in this system and observe state transfer between the two latent-symmetric network sites. This provides experimental confirmation of the latent symmetry present in the photonic network.

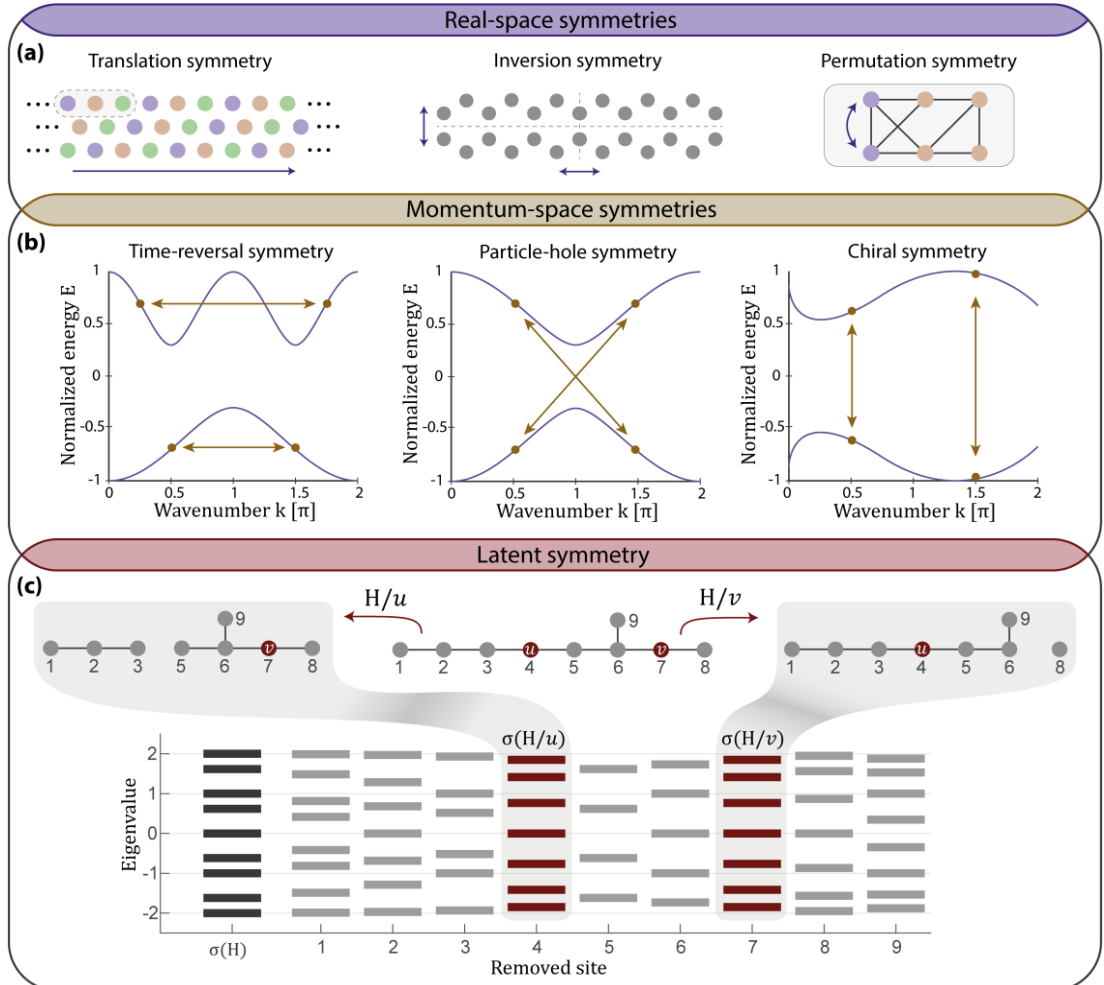

*Figure 1: Latent-symmetric networks (**a**) Symmetries such as translation and inversion symmetry, as well as permutation symmetry are defined in real space and visible with the bare eye. (**b**) The three symmetries time-reversal, particle-hole and chiral symmetry impose symmetry conditions on the energy bands, thus may considered momentum-space symmetries. (**c**) Latent or 'hidden' symmetries between two network sites u and v arise if the set of eigenvalues of the network with either one of these sites removed are equal despite a lack of a permutation symmetry between u and v. The two sites, u and v, are referred to as cospectral.*

We study latent symmetry in the network that is shown in Fig. 2a, which can be expressed by a graph containing nine sites, where the edges connecting them represent coupling. All network sites have equal on-site potentials and all couplings have the same value. Mathematically, any linear $N$-site network can be described by the tight-binding Hamiltonian

$$\mathbf{H} = \sum_i^N E_i |i\rangle\langle i| + \sum_{i,j}^N k_{i,j} |i\rangle\langle j|, \tag{1}$$

with $E_i$ the on-site potential on site $i$ and $k_{i,j}$ the coupling between two sites $i$ and $j$. Two sites $u$ and $v$ of a network **H** are latent symmetric if there is no permutation symmetry between them, but the two sites are *cospectral*. The cospectrality of two network sites is in turn defined by looking at the set of eigenvalues of the networks resulting from the removal of the distinct networks sites (Fig. 1c). If the networks that result from the removal of $u$ and $v$, respectively, share the same set of eigenvalues, $u$ and $v$ are cospectral. For the studied nine-site network, the latent symmetric sites are $u = 2$ and $v = 6$ (Fig. 2a). Along with the latent symmetric network sites, so-called singlet sites can be defined[23] as a site $w$ that has the same "distance" to the two cospectral sites $u$ and $v$, in our case $a = 4$, $b = 8$ and $c = 9$. At these sites, any connection may be attached to the network without breaking the latent symmetry of $u$ and $v$, which allows for arbitrary extensions of the initial network (Fig. 2b). Furthermore, the presence of singlet sites results in an interesting behaviour of double-site excitations on $u$ and $v$: An anti-symmetric input state, residing in $u$ and $v$ with opposite amplitude, will interfere destructively at all singlet sites[24]. Therefore, when the network is extended at those sites, the anti-symmetric state cannot leave the initial substructure.

The latent symmetry inherent in the studied network moreover has an additional impact on the dynamics of the system and can be leveraged for the transfer of states. The dynamics of a network **H**, hence the evolution of an arbitrary input state through the network, is given by the continuous-time quantum walk

$$|\psi(t)\rangle = \mathbf{U}(t)|\psi(t=0)\rangle, \tag{2}$$

with the time-evolution operator $\mathbf{U}(t) = \exp(-it\mathbf{H})$. The probability of an initial excitation on site $i$ being transferred via the network to site $j$ at a time $t$ is defined by the time-dependent transfer fidelity $F_{i,j}(t) = |U_{i,j}(t)|^2$ and quantifies the reliability of the state transfer. While various approaches to perfect state transfer (PST), which reaches a fidelity of unity at some finite time $t$, have been studied[25–31], they entail rather stringent constraints on the network that, in practice, prevent perfect transfer. Latent symmetries can be used to relax these conditions and to allow for greater flexibility in network design while nevertheless ensuring what has been termed *pretty good state transfer* (PGST)[32–40]. Crucially, while PGST is fundamentally precluded from achieving a perfect transfer efficiency, its fidelity in principle can approach unity as closely as desired within a finite time. In other words, for every $\varepsilon > 0$, there is a time $\tau$ such that $F_{i,j}(\tau) > 1 - \varepsilon$. Starting with a latent-symmetric network, its parameter space (on-site potentials and couplings) can be tuned in a way that it features PGST between the two latent-symmetric network sites, by meeting only two additional constraints related to the eigenvalues of $\mathbf{H}$[21]. Observation of PGST thus experimentally proves the existence of a latent symmetry inherent in the studied network. The numerically calculated fidelity $F_{u,v}$ for the network considered in this work is depicted in Fig. 2c over six orders of magnitude. It behaves as a quasi-random series of isolated fidelity peaks whose maximum approaches unity for large but finite times.

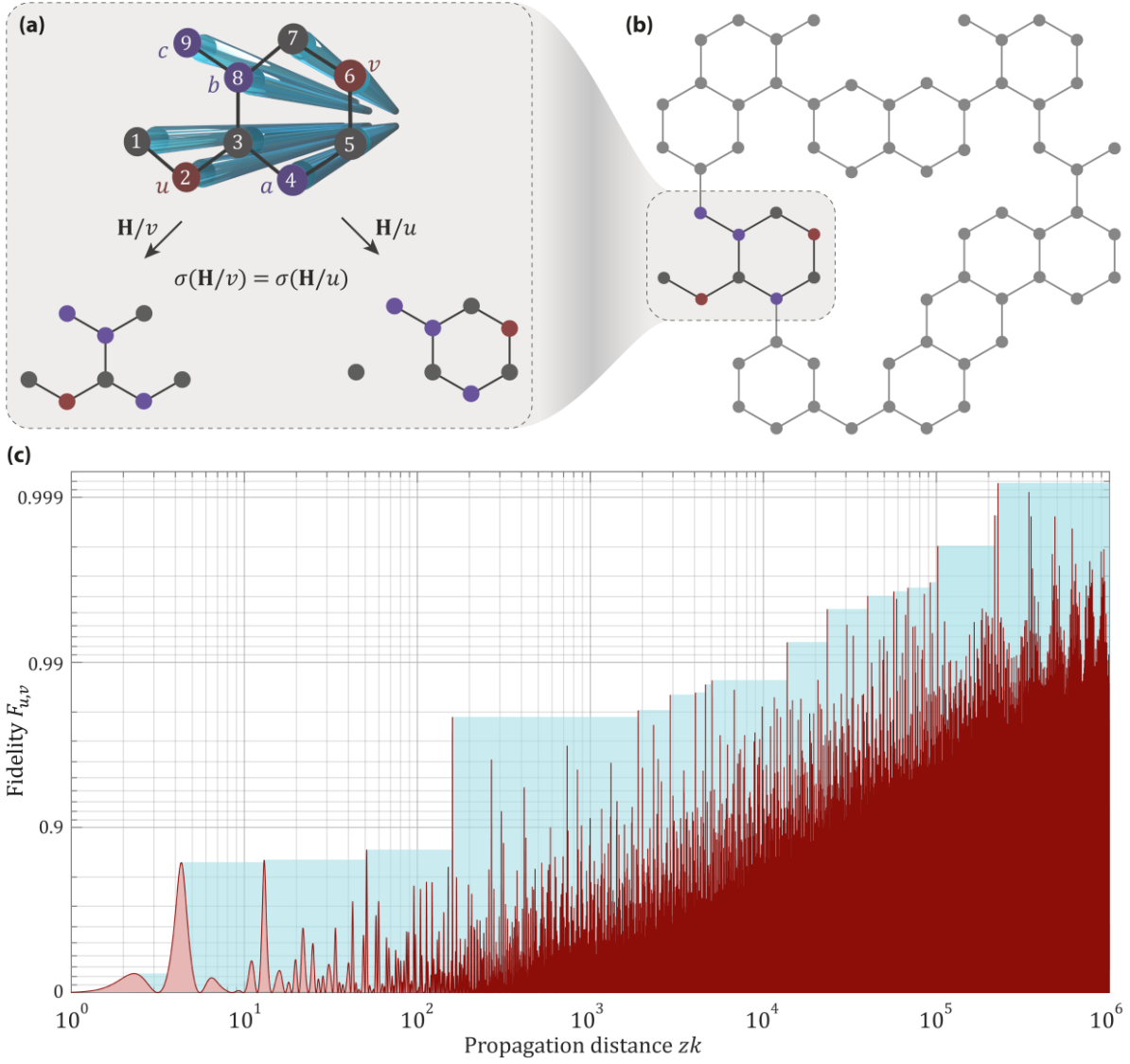

*Figure 2: State transfer in latent-symmetric networks (a) The implemented nine-site latent-symmetric network using evanescently coupled fs laser-written waveguides with uniform coupling. The latent symmetric sites $u = 2$ and $v = 6$ are marked red. Removing either site $u$ or site $v$ yields two graphs with the same spectrum (set of eigenvalues) $\sigma$. Thus, u and v are cospectral network sites (see Fig. 1c). Since there is no permutation symmetry between u and v, the two sites are latent symmetric. (b) Connecting arbitrary networks to the singlet sites $a = 4, b = 8$ and $c = 9$ does not break the symmetry between u and v. This makes it possible to extend the network indefinitely. (c) The simulated fidelity $F_{u,v}$ for large normalised propagation distances zk plotted logarithmically. The blue area in the background depicts the maximum value of F so far at this point. The maximum value of fidelity is approaching unity via a series of peaks with a maximum of $F_{u,v} \approx 0.999$ for propagation distances $zk \sim 10^5$.*

To experimentally implement the network containing a latent symmetry, we employ femtosecond laser-written waveguide arrays[41], an experimental platform that has previously been used to demonstrate perfect state transfer[42,43]. Our photonic system maps the time evolution of the network onto the propagation of light along the waveguide array. Every network site then corresponds to a waveguide in the experimental implementation, whereas the coupling between two sites is determined by the waveguides' geometric separation from one another and the angle between them, as well as the wavelength of the light. We design our system for operation at 814nm. Exciting either site $u = 2$ or site $v = 6$, we expect a transfer fidelity that approaches unity via a series of peaks as depicted in Fig. 2c. In our experiment, we observe the first of those peaks.

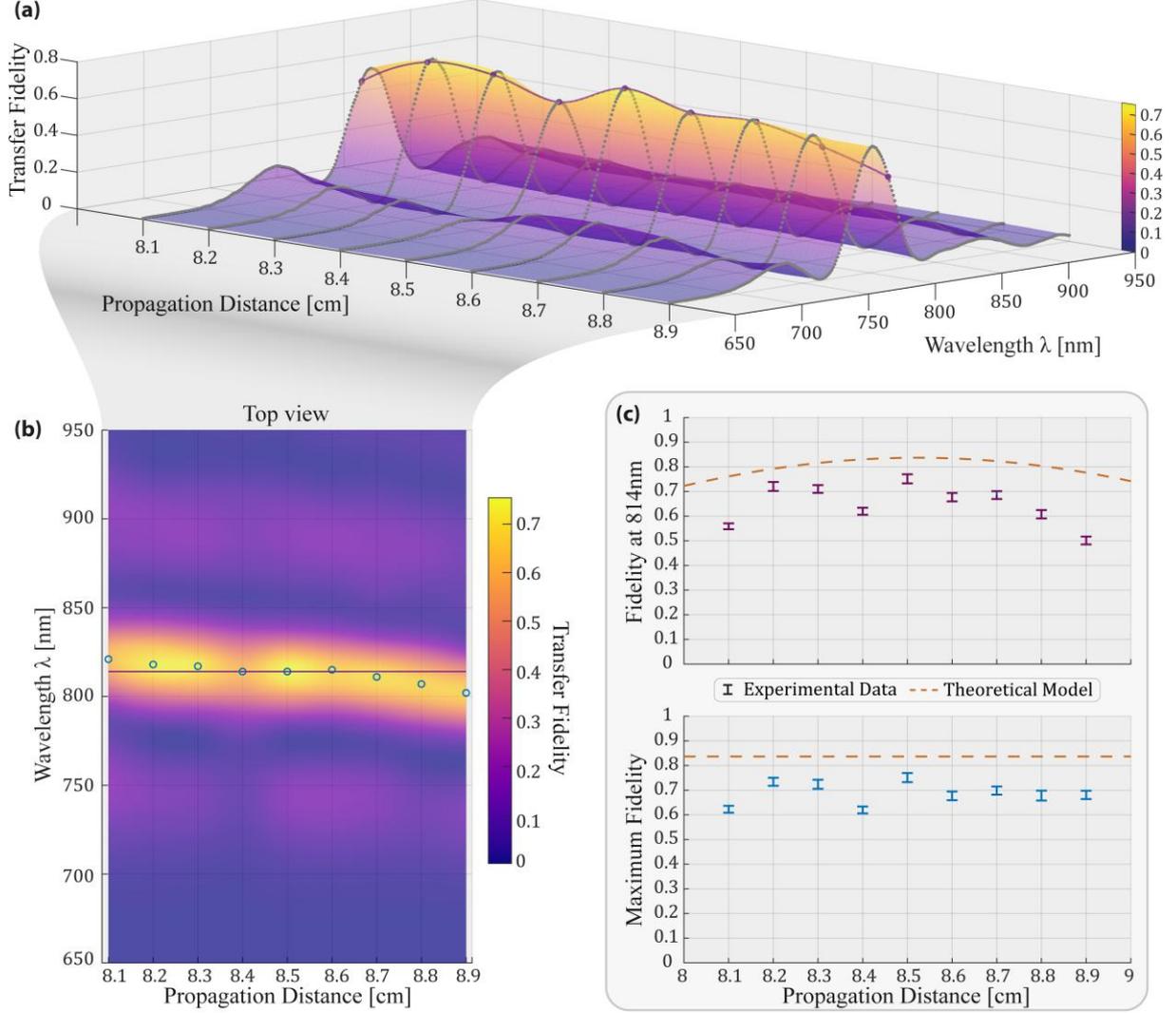

*Figure 3: Observation of classical state transfer (**a**) The experimentally measured fidelity of the state transfer from v to u as a function of the wavelength λ and different propagation distance z. Each propagation distance belongs to a distinct sample. The purple line connects data with the design wavelength of 814 nm. (**b**) Top view of the 3d-Plot in (a). Blue circles are the maximum of fidelity for each propagation distance. The wavelength at which the maximum occurs shifts to lower wavelengths with a change of propagation distance. (**c**) The measured fidelity for a fixed wavelength (top) and the maximum fidelity (bottom), as a function of the propagation distance. The error of transfer fidelity is dominated by the background noise.*

To explore the vicinity of the fidelity peak, we vary the effective propagation length, given by the product of propagation distance $z$ and coupling coefficient $k$. The length was increased in nine discrete steps, each of which was inscribed as a separate waveguide structure. The coupling $k$ is continuously adjustable via the probe wavelength, as the mode field and thus the coupling increases with wavelength.

We observe the output intensity distribution across the entire structure. The intensity evolution in waveguide $u$ resulting from a single-site excitation of waveguide $v$ directly corresponds to the fidelity of state transfer $F_{u,v}(z) = |U_{u,v}(z)|^2$. The results presented in Fig. 3 are confirmed by the behaviour in simulations based on the ideal design parameters. A maximum fidelity of $F_{u,v} = 75\%$ is achieved for a propagation distance of $l_k = 8.5$cm. Fig. 3b clearly demonstrates that increasing the propagation distance leads to a shift of the maximum fidelity towards shorter wavelengths, in line with the decreasing mode field diameters and coupling coefficients. Note that the peculiar nature of PGST provides the beneficial character that it will inevitably approach $F_{u,v} = 1$ at longer propagation lengths (see Fig. 2c).

In the final step, we experimentally demonstrate quantum state transfer of two photons in the network and confirm the behaviour of an anti-symmetric input state. To this end, we use an anti-symmetric two-photon excitation of the two cospectral sites and observe the two-photon correlations at the output of the latent-symmetric network. This state is prepared by means of a balanced beam splitter and a $\pi/2$-phase shift prior to the latent-symmetric network. For indistinguishable photons, this gives the anti-symmetric state $|\psi_{\text{in}}\rangle = (|2_u, 0_v\rangle - |0_u, 2_v\rangle)/\sqrt{2}$. Distinguishable photons (photons that arrive delayed at the beam splitter) serve as a reference. The theoretically and experimentally obtained correlations, are presented in Fig. 4a and 4b. We observe suppression of coincidences between singlet sites $a = 4$, $b = 8$ and $c = 9$, as a result of destructive interference of the anti-symmetric input state.

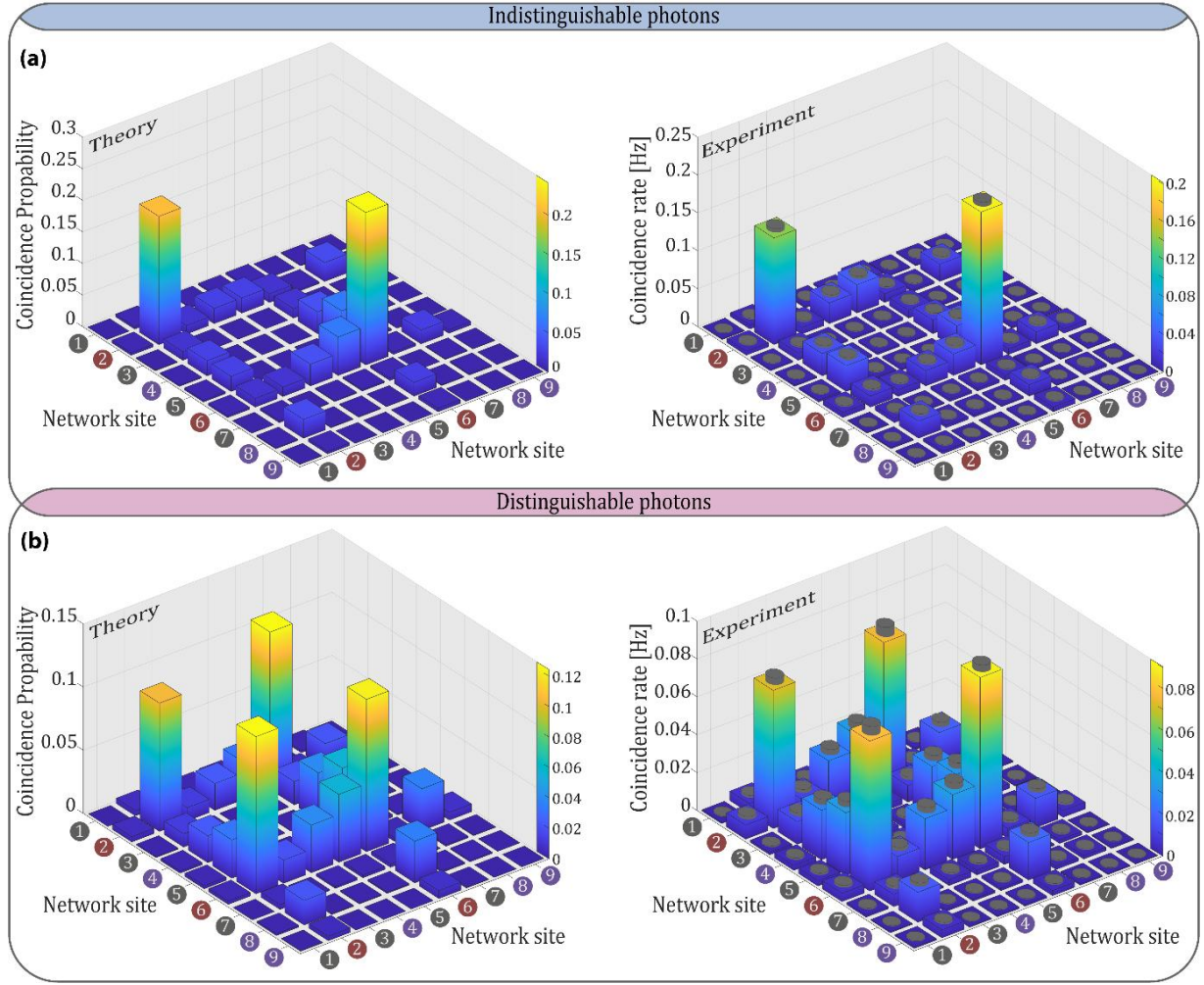

*Figure 4: The simulated and measured coincidence matrices for an anti-symmetric input state for (a) indistinguishable photons and (b) distinguishable photons. The network transfers the input state, resulting on indistinguishable photons leaving the network bunched either at site $u$ or $v$, while for distinguishable photon coincidences between the two sites $u = 2$ & $v = 6$ are measured. There are no correlations between the singlet sites $a = 4, b = 8$ and $c = 9$ in both cases. Errors are depicted by grey cylinders and are dominated by the statistical Poisson error.*

In our work, we design a nine-site latent symmetric network, which is implemented using integrated photonics. We observe single-site excitation state transfer with a maximum transfer fidelity of $F = 75\%$ and measure two-photon state transfer for indistinguishable and distinguishable photons. This experimentally proves the existence of a latent symmetry inherent to the implemented network. Compared to traditional perfect state transfer, this approach enables greater flexibility in designing networks that support efficient quantum state transfers, resulting in a larger pool of potentially usable networks. Our implementation of latent symmetries paves the way to design and realize more functionalities based on latent symmetry, such as systems that exchange energy with the environment[18], or topological states protected by latent symmetry[19]. In this vein, we foresee latent symmetries to potentially play a pivotal role in quantum cryptography, in particular when transferring states in a secure manner.

### Acknowledgments

We thank C. Otto for preparing the high-quality fused silica samples used for the inscription of all photonic structures employed in this work. AS acknowledges funding from the Deutsche Forschungsgemeinschaft (grants SZ 276/9-2, SZ 276/19-1, SZ 276/20-1, SZ 276/21-1, SZ 276/27-1, and GRK 2676/1-2023 'Imaging of Quantum Systems', project no. 437567992). AS also acknowledges funding from the Krupp von Bohlen and Halbach Foundation as well as from the FET Open Grant EPIQUS (grant no. 899368) within the framework of the European H2020 programme for Excellent Science. AS and MH acknowledge funding from the Deutsche Forschungsgemeinschaft via SFB 1477 'Light–Matter Interactions at Interfaces' (project no. 441234705). TAWW is supported by a European Commission Marie Skłodowska-Curie Actions Individual Fellowship (project no. 895254).

## Sample fabrication and characterisation

We inscribed our coupled wave guide systems using the femtosecond laser direct writing[1]. Ultrashort laser pulses of 270 fs duration from a frequency-doubled fibre laser system (Coherent Monaco) at a wavelength of 517 nm and a repetition rate of 333 kHz were focused into a 150 mm × 25 mm × 1 mm fused silica chip (Corning 7980) by means of a microscope objective (×50, numerical aperture = 0.6). The sample was positioned with 50 nm precision by a three-axis motorized translation stage (Aerotech ALS180).

The waveguide array was designed with an angle between the network sites of 30° to minimize undesired not nearest-neighbour coupling. To ensure that the diagonal and vertical couplings are the same, coupling characterisation scans were performed beforehand. The diagonal and vertical distances were chosen to be $x_{\text{diag}} = 23.35\,\mu\text{m}$ and $x_{\text{vert}} = 23.1\,\mu\text{m}$, respectively. Waveguides $u$ and $v$ were connected to the front facet using a fanning section, to match the distance of the fibres in the incoupling fibre array (82 μm). For the fanning, $\cos^2$-trajectories with small changes in the $x$ and $y$ direction relative to the $z$ direction were used to achieve minimal losses due to waveguide bending. The structure ended at the end facet without fanning and it was coupled to the detectors in free-space.

The network was then characterized using a supercontinuum white light source (*SuperK EXTREME*) and a monochromator with a bandwidth of 2.5nm (*LLTF Contrast*) as an input for the single site excitation coupled into the waveguide via a microscope objective (×10, numerical aperture = 0.2). The wavelength of the coupled light was changed by steps of 1nm. As the wavelength increases, the mode field of the propagating light in the waveguide expands. The waveguides' coupling strength is dependent on the overlap of the mode fields and, consequently, changing the wavelength alters the coupling strength. An increase of the mode field yields an increase of the coupling strength. Note that the dependence of wavelength and coupling strength is not linear. The output intensity of each waveguide was measured with a CMOS camera (Basler ace) and normalized to the total output intensity. The fidelity of state transfer from $v$ to $u$ and vice versa is then given by

$$F_{u,v} = \frac{I_v - I_{\text{bg}}}{I_{\text{tot}}}, \qquad F_{v,u} = \frac{I_u - I_{\text{bg}}}{I_{\text{tot}}} \tag{1}$$

where $I_{\text{tot}} = \sum_i^9 (I_i - I_{\text{bg}})$ is the total intensity in all sites. The background intensity $I_{\text{bg}}$ was removed for every site and also determines the maximum error of intensity.

For the two-photon measurement horizontally polarized wavelength-degenerate photon pairs at 814 nm are used, which are generated by type-I spontaneous parametric down-conversion from a continuous-wave pump at 407 nm in a bismuth borate crystal. To achieve the anti-symmetric input state, a second chip with a balanced beam splitter was fabricated and coupled to the latent-symmetric network. The output state of the beam splitter was characterized via a HOM-Dip measurement. By changing the coupling angle between the BS chip and the PGST chip, it is possible to increase the optical path length of one input site relative to the other one. This way the phase between the two inputs could be adjusted to get a total phase shift of $\varphi_{\text{tot}} = \pi$ for the anti-symmetric input state. The outputs were split by a beam splitter and imaged (objective) onto multimode fibres, which are connected to avalanche-photodiodes and a correlation card (*TimeTagger* by *Swabian Instruments*). The difference in detection and incoupling efficiency could be corrected afterwards by normalization over the total counts in each detector. The coincidences at the diagonal were measured 7200s, the coincidences at the off-diagonals were measured 3600s. Since the matrix is symmetric and $C_{ij} = C_{ji}$ without detection differences, each possible output was measured the same time. The measurement was performed for the latent-symmetric network with propagation distance $z = 8.6$ cm.

## Network design

Applying the algorithm presented by Röntgen et al.[2], the Hamiltonian implemented in this work could be designed. As desired, the design process started with a network having two cospectral sites as shown in Fig. 2a of the main text. Its parameter space $\zeta = \{k, E_4, E_9\}$ is defined by the coupling strength $k$ and the on-site potentials $E_4 = E_9 = E$ on sites 4 and 9. The on-site potential of the remaining sites is set to zero, as this is the smallest parameter space for which PGST is possible. With these values, the isospectral reduction[3], which reduces the dimension of the Hamiltonian, while keeping (nearly) all of its spectral information, $\mathbf{R}_S(\mathbf{H}, \lambda) = \mathbf{H}_{SS} - \mathbf{H}_{S\bar{S}}(\mathbf{H}_{\bar{S}\bar{S}} - \mathbf{I}\lambda)\mathbf{H}_{\bar{S}S}$ over the subset $S = \{u, v\}$ of $\mathbf{H}(\zeta)$ yields

$$A(\xi, \lambda) = \frac{k^2(E^2(2\lambda^3 - 3k^2\lambda) + E(-4k^4 - 4\lambda^4 + 11\lambda^2 k^2) + 2\lambda^5 + 7k^4\lambda - 8k^2\lambda^3)}{\lambda(E^2(\lambda^3 - 2k^2\lambda) + E(-3k^4 - 2\lambda^4 + 7\lambda^2 k^2) + \lambda^5 + 5k^4\lambda - 5k^2\lambda^3)} \quad (2)1$$

$$B(\xi, \lambda) = \frac{k^4(E^2\lambda - 3E\lambda^2 + 2Ek^2 + 2\lambda^3 - 3k^2\lambda)}{\lambda(E^2(\lambda^3 - 2k^2\lambda) + E(-3k^4 - 2\lambda^4 + 7\lambda^2 k^2) + \lambda^5 + 5k^4\lambda - 5k^2\lambda^3)} \quad (3)2$$

with $A(\zeta, \lambda), B(\zeta, \lambda)$ the diagonal and off-diagonal entries and $\lambda$ the non-linear eigenvalue of the reduced $2 \times 2$ matrix. We will now proceed through the steps outlined by Röntgen et al.[2], beginning with a check for strong cospectrality of $u$ and $v$. As theorem 3.8 in the work of Kempton et al.[4] states, two vertices $u$ and $v$ are *strongly cospectral* if they are cospectral, and, additionally, all non-linear eigenvalues $\lambda$ of the isospectral reduction are simple, which applies if all roots of the corresponding characteristic polynomial are simple. As outlined in Ref. 2, the polynomial can be decomposed into three parts, $P = P_+ \cdot P_- \cdot P_0$. Here, $P_\pm$ belong to the eigenvalues whose eigenmodes have positive (negative) parity on $u$ and $v$ and $P_0$ belongs to the eigenvalues whose eigenmodes vanish on $u$ and $v$. Under mild conditions (see Sec. III, point (3) of Ref. 2), which apply for our choice of $k, E$ (see below) the polynomials $P_\pm$ are given by the numerators of the rational functions $p_\pm/q_\pm = A(\xi, \lambda) \pm B(\xi, \lambda) - \lambda$. We note that one has to first cancel any common factor of $p_\pm$ and $q_\pm$, and also expand them afterwards such that the leading-order coefficient of the resulting polynomials $P_\pm$ is equal to either $+1$ or $-1$.

If, for a subspace $\zeta' \subseteq \zeta$, $P_\pm(\zeta', \lambda)$ individually only have simple roots and no common roots, $u$ and $v$ are strongly cospectral. This yields the restricted parameter space $\zeta'=\{k, E \neq \pm k(1 \pm \sqrt{2})\}$.

To achieve PGST between the sites $u$ and $v$, $P_\pm$ have to be irreducible and additionally fulfill[5]

$$\frac{\text{Tr}(P_+)}{\deg(P_+)} \neq \frac{\text{Tr}(P_-)}{\deg(P_-)}. \quad (4)$$

(4)In our experiment, we have $E = 0$, and $k \approx \frac{54}{100}$. This yields $P_- = \lambda^2 - \frac{729}{1250}$, $P_+ = \lambda^6 - \frac{5103\,\lambda^4}{2500} + \frac{5845851\,\lambda^2}{6250000} - \frac{387420489}{3906250000}$, and $P_0 = \lambda$. These polynomials fulfill all necessary conditions, so that the Hamiltonian supports PGST.

## Two-Photon state preparation and theory calculation

Using the transition probabilities $P_{D,B}(\mathbf{r},\mathbf{s};\mathbf{U})$ from an input state described by the input *mode occupation list* $\mathbf{r} = (r_1, \ldots, r_n)$ to an output state described by the *output* mode occupation list $\mathbf{s} = (s_1, \ldots, s_n)$, the correlation matrices for distinguishable and indistinguishable photons could be calculated. For each input/output mode occupation list $\mathbf{q}$ a *mode assignment list*

$$\mathbf{d}(\mathbf{q}) = \oplus_{j=1}^{n} \oplus_{k=1}^{n} (j) = (\underbrace{1,\ldots,1}_{q_1}, \underbrace{2,\ldots,2}_{q_2}, \ldots, \underbrace{n,\ldots,n}_{q_n}) \tag{3}$$

can be defined. For distinguishable particles we then get[6]

$$P_D(\mathbf{r},\mathbf{s};\mathbf{U}) = \frac{1}{\prod_{j=1}^{n} s_j!} \operatorname{perm}(|\mathbf{M}|^2), \tag{4}$$

for indistinguishable bosons

$$P_B(\mathbf{r},\mathbf{s};\mathbf{U}) = \frac{1}{\prod_{j=1}^{n} r_j! s_j!} |\operatorname{perm}(\mathbf{M})|^2. \tag{5}$$

The matrix $\mathbf{M}$ is defined as

$$M_{j,k} = U_{d_j(\mathbf{r}), d_k(\mathbf{s})} \tag{6}$$

and $\operatorname{perm}(\mathbf{M})$ is the permanent of $\mathbf{M}$.

The correlation matrix is solely defined by the permanents

$$\Gamma_D(\mathbf{r},\mathbf{s};\mathbf{U}) = \operatorname{perm}(|\mathbf{M}|^2), \qquad \Gamma_B(\mathbf{r},\mathbf{s};\mathbf{U}) = |\operatorname{perm}(\mathbf{M})|^2.$$

Launching a two-photon state $|\psi_{\text{in}}\rangle = |1,1\rangle$ into a balanced beam splitter described by

$$\mathbf{U}_{\text{BS}} = \frac{1}{\sqrt{2}} \begin{pmatrix} 1 & i \\ i & 1 \end{pmatrix} \tag{7}$$

yields the input mode occupation list $\mathbf{r} = (1,1)$ and the input mode assignment list $\mathbf{d}(\mathbf{r}) = (1,2)$. Considering the probabilities for the possible output states, we see the expected photon bunching (HOM-dip): $P_B(\mathbf{r},\mathbf{s}=(2,0);\mathbf{U}_{\text{BS}}) = 0.5, P_B(\mathbf{r},\mathbf{s}=(0,2);\mathbf{U}_{\text{BS}}) = 0.5, P_B(\mathbf{r},\mathbf{s}=(1,1);\mathbf{U}_{\text{BS}}) = 0$.

This corresponds to the superposition state

$$|\psi_{\text{out}}\rangle = \frac{|2,0\rangle + |0,2\rangle}{\sqrt{2}}. \tag{8}$$

Applying a phase shift of $\pi/2$ between the two outputs, by changing the optical path length of one output, generates an anti-symmetric state

$$|\psi_{\text{out}}\rangle = \frac{|2,0\rangle - |0,2\rangle}{\sqrt{2}}. \tag{9}$$

This state was used as an input state for the latent symmetric network. The output probabilities $P_{B/D}$ could be calculated by considering the total time evolution operator

$$\mathbf{U}_{\text{tot}} = \mathbf{U}_{\text{PGST}} \mathbf{U}_\varphi \mathbf{U}_{\text{BS}}, \tag{10}$$

and the input mode assignment list $d(r) = (2,6)$, where $\mathbf{U}_{\text{BS}}$ only acts as a balanced beam splitter on sites 2 and 6 and $\mathbf{U}_{\varphi(6,6)} = e^{i\pi/2}$ acts only on site 6. To compare these probabilities to the experimentally obtained data, the off-diagonal terms had to be divided by two, due to the symmetry of the coincidence matrix.